\documentstyle{mn}
\begin{document}

\title{Gravothermal Catastrophe in Anisotropic Spherical Systems}

\author[M.Magliocchetti, G.Pucacco, E.Vesperini]{M. Magliocchetti$^{1,3}$,
       G. Pucacco$^{2,3}$,
      E. Vesperini$^{4}$\\
$^1$ Institute of Astronomy, Madingley Road, Cambridge
   CB3 0HA, England\\
$^2$ Department of Physics, Universit\`a di Roma ``Tor Vergata'', 
          Via della Ricerca Scientifica 1, I-00133 Roma\\
$^3$ I.C.R.A.-International Centre for Relativistic
   Astrophysics, P.le Aldo Moro 2, I-00185 Roma\\ 
$^4$ Department of Physics and Astronomy, University of Massachusetts,
Amherst, 01003, MA, USA}

\maketitle

\begin{abstract}
In this paper we investigate the gravothermal instability of spherical
stellar systems endowed with a radially anisotropic velocity
distribution. We focus our attention on the effects of anisotropy on
the conditions for the onset of the instability and in particular we
study the dependence of the spatial structure of critical models on
the amount of anisotropy present in a system. The
investigation has been carried out by the method of linear series
which has already been used in the past to study the gravothermal
instability of isotropic systems. 

We consider models described by King, Wilson and Woolley-Dickens
distribution functions. In the case of King and
Woolley-Dickens  models, our results show that, for quite  a wide range of
amount of anisotropy in the system, the critical value of the concentration of
the system (defined as the ratio of the tidal to the King core radius
of the system) is approximately constant and equal to the
corresponding value for isotropic systems. Only for very anisotropic
systems the critical value of the concentration starts to change and
it decreases significantly as the anisotropy increases and penetrates
the inner parts of the system. For Wilson models the decrease of the 
concentration of critical models is preceded by an intermediate regime
in which critical concentration increases, it reaches a maximum and then it
starts to decrease. The critical value of
the central potential  always decreases as the anisotropy increases. 

\vskip\baselineskip
\end{abstract}


\section{Introduction}

It is well known that the phenomenon of gravothermal instability is 
strictly related to the property of self-gravitating systems of having 
a negative specific heat. In 1962 Antonov showed that there is no  state  
characterized by a local maximum in the entropy for stellar systems with  
fixed mass and energy, contained in a spherical box with radius greater  
than the critical value $r_{\rm crit}=-0.335({GM^2}/{E})$ where $M$ is the total 
mass of the system and $E$ its total energy.  Lynden-Bell and Wood (1968)
interpreted this  result in their pioneering work on the stability of isotropic
isothermal stellar systems truncated in  radius, establishing the conditions for
the onset of the phenomenon, called then {\it gravothermal catastrophe}.
 The investigation was carried out by means of the 
{\it method of  linear series} formulated for the first time by
Poincar\'e (1885); this  technique
is connected with the study of global stability of the equilibria of a
generic system. 

In the case of isothermal configurations, the
equilibrium states in the  presence of encounters are determined by finding the
conditions for a maximum of the total entropy $S$, keeping fixed the total
energy and mass, $E$ and $M$, of the system. The phase-space distribution
function that  describes a system under these requirements turns out to be the
Maxwell-Boltzmann function,

\begin{eqnarray}
f(\varepsilon)=A\:e^{-\beta\varepsilon},
\label{MB} \end{eqnarray}
where 

\begin{eqnarray}
\varepsilon={1 \over 2} {v^2} + \Phi (r) 
\end{eqnarray}
is the energy per unit mass  of  single stars, $\beta$ is the inverse of the
``temperature", $A$ is a normalization constant and $\Phi(r)$ is the
self-consistent gravitational potential obtained by solving the Poisson
equation.  Lynden-Bell and Wood focussed their attention on the study of  
the stability of a system against perturbations keeping 
constant, besides the total mass and energy, the radius of the system also.
The critical value of the concentration of the system, as measured by
the ratio of the central density, $\rho_{\rm c}$, to the average density inside
the boundary of the system, $\rho_{\rm b}$, is equal to $\rho_{\rm c} /
\rho_{\rm b}=709$ which coincides with the result found by Antonov.

An analogous investigation was later carried out by Katz (1978, 1980) for
more realistic models of stellar clusters. Katz 
studied the gravothermal instability of  King (1965),
Woolley-Dickens (1961) and Wilson (1975) models, which are all
characterized by distribution functions with a cut-off in the energy space. In
this case,  the study of the stability was realized keeping fixed the 
total mass, the total energy and the multiplicative normalization
parameter $A$ which reflects the indetermination in the zero of the chemical
potential (see the appendix of Katz (1980) for the discussion on the motivation 
for this choice). In these cases, the onset of gravothermal instability 
occurs for  the following values of the central dimensionless
relative potential, 
$\tilde \Psi_{\rm c} \equiv \beta \Psi_{\rm c} $, 
\begin{eqnarray}
\begin{array}{lll}
1)\;\tilde \Psi_{\rm c}=7.65,&\hspace{.5cm}& Woolley-Dickens\; models, \\
2)\;\tilde \Psi_{\rm c}=7.40,&\hspace{.5cm}& King\; models, \\
3)\;\tilde \Psi_{\rm c}=6.67,&\hspace{.5cm}& Wilson\; models.
\end{array}
\label{cc} \end{eqnarray}

Taking anisotropy in the velocity
distribution into account is an essential step in order to improve the
reliability of energy-truncated models to better
describe the  structure and evolution of clusters. The important role played by
anisotropy has been recognized since the dawn of
the study of globular clusters (Woolley \&  Robertson, 1956). The main goal of
the present paper is that of extending the study of gravothermal instability to 
spherical systems endowed with an  anisotropic velocity distribution and
determining the dependence of the critical value of $\tilde \Psi_{\rm c}$ on the
amount of anisotropy present in the systems.  We point out that our analysis is
meant to provide only a general indication of the role of radial anisotropy on
the onset of gravothermal instability, providing a simple and
straightforward estimate of its effect as measured by the critical
value of the central potential and of the concentration of the system
(given by the ratio of the total radius of the system to the
King core radius). A much deeper insight in the collisional
evolution of anisotropic systems can be obtained only by other methods
of investigation such as $N$-body simulations, integration of the
Fokker-Planck equation and gaseous models (see e.g. Takahashi 1995,
1996, Takahashi, Lee \& Inagaki 1997, Spurzem \& Takahashi 1995, Giersz \&
Spurzem 1994, Bettwieser \& Spurzem 1986, Giersz \& Heggie 1994a,b, 1996, 1997,
Louis \& Spurzem 1991, Spurzem 1991, Spurzem \& Aarseth 1996, 
for some interesting and detailed investigations on the evolution of
anisotropic stellar systems). In Sects. 3.3 and 4, we compare the global
indications resulting from the present study with the specific findings of some
of the most recent works cited above. 

The scheme of the paper is the following:
in Section 2 we summarize the main properties of these models, in Section 3 we
study their stability by means of the method of linear series discussing the
results and, finally, in Section 4 we summarize the main conclusions.  

\section{Anisotropic models}

The relevance of an anisotropic model for the description of real stellar 
systems arises when studying the first periods of life of stellar systems:
it is possible to demonstrate that, at the end of the process 
of the  violent relaxation (Lynden-Bell, 1967), stellar systems are
anisotropic because they still have stars  mainly distributed on
radial orbits,   
and, although they tend to relax and become isotropic on a 
time scale which is of the order of the binary relaxation time (see, e.g., 
Binney \& Tremaine, 1987), if the conditions for the onset of the gravothermal
instability take place before their total isotropization, anisotropy cannot
be left out of consideration. It must be remarked that, in evolving
systems like globular clusters, anisotropy is not simply removed by relaxation.
Rather, it is even increased in the halo, due to the high energy stars coming
from the strong relaxing core (Takahashi 1996) and it is also expected to
gradually penetrate into the inner regions as the core collapses. The
situation is different for tidally truncated systems in which stars
can escape at a significant rate from the cluster: in this case it has
been shown (Giersz \& Heggie 1997, Takahashi et al. 1997)
that  anisotropy production can be partially or completely 
balanced by the evaporation of stars which would occur preferentially
for stars on radial orbits.

In any case, if primordial anisotropy changes the conditions for the onset of
gravothermal instability, this can also have important consequences on the
subsequent evolution of anisotropy itself. In the present Section we describe
the properties of the anisotropic variants of the energy-truncated models
mentioned above, which, in the subsequent one, will be submitted to the
analysis of gravothermal stability.

\subsection{Description of anisotropic spherical models}

Following arguments similar to those of  Tremaine (1986) in the discussion of
the process of violent relaxation, the distribution functions introduced
to describe anisotropic systems are obtained by multiplying by a
factor of the form

\begin{eqnarray}
k(L^2)\equiv e^{-\frac{L^2}{2\sigma^2 r_{\rm a}^2}}
\label{K} \end{eqnarray} 
the isotropic King, Woolley-Dickens and Wilson distribution functions, which
therefore can be written in the form

\begin{eqnarray}
\begin{array}{lll}
1)\;f_{WD}(\epsilon,L^2)=A\:e^{-\frac
{\epsilon}{\sigma^2}}
\:e^{-\frac{L^2}{2\sigma^2r_{\rm a}^2}},\\
2)\;f_{K}(\epsilon,L^2)=A\:\left(e^{-\frac
{\epsilon}{\sigma^2}}-1\right)\:e^{-\frac{L^2}{2\sigma^2r_{\rm a}^2}},\\
3)\;f_{W}(\epsilon,L^2)=A\:\left(e^{-\frac
{\epsilon}{\sigma^2}}
+\frac{\epsilon}{\sigma^2}-1\right)\:e^{-\frac{L^2}{2\sigma^2r_{\rm a}^2}}.
\end{array}
\label{DF} \end{eqnarray}
The above definitions hold for $\epsilon<0$, while it is assumed that 
$f(\epsilon)=0$ for $\epsilon>0$. In each distribution 
function, the parameter $\beta$ has been substituted with the inverse of the
isothermal velocity scale $\sigma^2$. The dependence of the functions on the
square of the angular momentum
$L$  ensures the spherical symmetry of the global properties of the models. In
(\ref{K}), $r_{\rm a}$ is the {\it anisotropy radius} that represents the value of the
radius beyond which the orbits start to be mainly radial. The distribution
function $ f_{K}(\epsilon,L^2) $ is usually known as {\it Michie} model (Binney
\& Tremaine, 1987), whereas $ f_{W}(\epsilon,L^2) $ is simply the spherically
symmetric version of the generic function introduced by Wilson (1975) himself. 

In order to perform the integration of these models, we define the following 
dimensionless quantities:

\begin{eqnarray}
x\equiv\frac{r}{r_K},\hspace{1cm}\tilde{v}\equiv\frac{v}{\sigma},\label{adim1}
\end{eqnarray}
so that

\begin{eqnarray}
\tilde{\epsilon}\equiv\frac
{\epsilon}{\sigma^2}={1 \over 2} {\tilde v}^2 + \tilde{\Phi},
\hspace{.5cm}\tilde{\Phi}\equiv\frac{\Phi}{\sigma^2}.
\label{adim2} \end{eqnarray}
In (\ref{adim1}), 
\begin{eqnarray}
r_{\rm K}=\frac{3\sigma}{\sqrt{4 \pi G \rho_{\rm c}}}
\end{eqnarray}
is the {\it King radius} and gives the length scale. Calling $ r_{\rm t} $ the
{\it tidal radius} of the system, so that $ x_{\rm t} = r_{\rm t} / r_{\rm K} $,
the self-consistent relative potential 
\begin{eqnarray}
\tilde{\Psi}(x)=-\tilde{\Phi}(x)+\tilde{\Phi}(x_{\rm t})
\end{eqnarray}
and, from it, the physical properties of the models, is obtained
by solving the Poisson equation which can be written in terms of the
dimensionless quantities introduced in (\ref{adim1}) and (\ref{adim2}) as:

\begin{eqnarray}
\frac{1}{x^2}\:\frac{d}{dx}\:\left(x^2\:\frac{d\tilde{\Psi}}{dx}\right)=
\frac{d^2\tilde{\Psi}}{dx^2}+\frac{2}{x}\:\frac{d\tilde{\Psi}}{dx}=\nonumber\\
=\frac{4 
\pi G r_{\rm K}^2}{\sigma^2}\rho(\tilde{\Psi}(x),x))=
-9\:\frac{\rho(\tilde{\Psi}(x),
x)}{\rho_{\rm c}}
\label{poisson} \end{eqnarray}
where the last equality comes from the substitution of $r_{\rm K}$ with its 
explicit expression. The related Cauchy's problem is then solved with initial 
conditions $\tilde{\Psi}(0)=\tilde{\Psi}_{\rm c}$ and $\tilde{\Psi}'(0)=0$.

The degree of anisotropy of a given model is parametrized by the
dimensionless anisotropy radius
\begin{eqnarray}
\gamma\equiv\frac{r_{\rm a}}{r_{\rm K}}.
\end{eqnarray}
We have carried out the integration of anisotropic
Woolley-Dickens, King and Wilson  models by selecting different values of the
anisotropy parameter $\gamma$ and  of the central potential $\tilde \Psi_{\rm
c}$. $\gamma$ has been chosen to range  from almost isotropic configurations
($\gamma=100$) up to highly anisotropic  models ($\gamma=0.3$). Models with
higher values of $\tilde \Psi_{\rm c}$ tend 
all to the singular isothermal sphere; however, highly concentrated models,
pose a certain constraint on the range allowed to the anisotropy parameter. On
the other hand, following arguments analogous to those in Merritt et al. (1989),
it can be shown that singular models with a completely radial distribution
display unphysical behaviour.

\begin{figure}
\vspace{9cm}  
\includegraphics{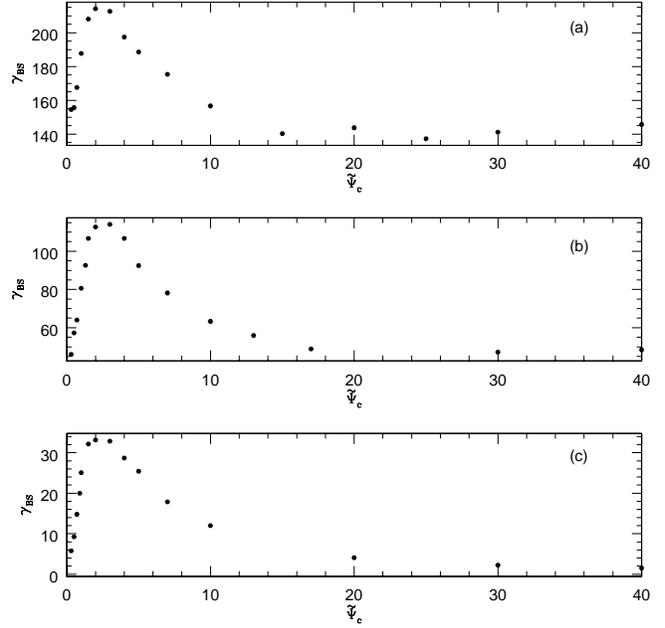} 
\caption{$\gamma_{\rm BS}$ versus the dimensionless central potential $\tilde
\Psi_{\rm c}$   for    
the first unbounded Woolley-Dickens (a), King (b) and
 Wilson model (c).
\label{fig:fig1} }
\end{figure}
\noindent

\subsection{Infinite models}

An  interesting feature of these models is the existence of definite pairs of
values ($\gamma,\tilde \Psi_{\rm c}$) that  identify unbounded models, i.e. 
models with an infinite  radius. In fact, there is a critical value of the
central potential, depending on  the degree of anisotropy characterising each
model, beyond which we have $r_{\rm t} \rightarrow \infty$ for every $\gamma$;
this  happens even though the distribution functions used for the description of 
these models are energy-truncated and the corresponding isotropic
configurations are always  bounded. Such remarkable peculiarity is
common to all anisotropic models, regardless of the form of the dependence
on the energy of the functions (\ref{DF}).

Following Stiavelli \& Bertin (1985) we consider the parameter
\begin{eqnarray}
\gamma_{\rm BS} \equiv \frac{1}{4\pi G A \sigma r_{\rm a}^2}.
\end{eqnarray}  
This parameter is related to our anisotropy parameter through the
following relation: 
\begin{eqnarray}
\gamma_{\rm BS}=\frac{\rho_{\rm c}}{9 \sigma^3A\gamma^2}.
\end{eqnarray}  
 
In Fig. 1 we have plotted $\gamma_{\rm BS}$ versus $\tilde {\Psi}_{\rm c} $
for the anisotropic
Woolley-Dickens, King and Wilson models having an infinite radius. These models
are singled out identifying, along the sequence at given $\tilde{\Psi}_{\rm c}$
parametrized by $ \gamma_{\rm BS} $, the first system for which the relative
potential goes as $ 1 / x $ at large radius. 

It is evident
that the behaviour is the same of that shown in the paper by Stiavelli \& Bertin
(1985) and thus it seems  that {\it this characteristic is common to all the 
spherical anisotropic models  having  a  dependence on the
angular momentum of the kind $e^{-L^2}$, regardless to the form of the
energy distribution}. 

Note that not all these unbounded models are relevant to model
real physical systems; in fact, while the total dimensionless mass of infinite  
Wilson and Bertin-Stiavelli models is finite, it diverges for 
anisotropic Woolley-Dickens and King models, making the latter models useful
only for those values of $\tilde {\Psi}_{\rm c}$ and $\gamma_{\rm BS}$ which
lead to a finite-size system.

\section{Gravothermal instability of spherical anisotropic models}

\subsection{Linear series of equilibrium}

The investigation of the gravothermal stability of the models
discussed in Section 2 has been  carried out by means of the 
method of linear series. This approach, introduced by Poincar\'e in 1885, is
based on the study of global stability of the equilibrium states of a generic
system.  In many stationary  problems, the description of a system depends on
some parameter other than the generalized coordinates $q_i$ ($i=1, \dots , n$) in
terms of  which we describe the configuration of the system. If we call $\mu$
this parameter and $U$  the potential energy of the system, the 
condition for the existence of an equilibrium is that the tangent plane to the 
$n$-dimensional surface 
\begin{eqnarray}
U(\mu,q_i)=\rm const
\end{eqnarray} 
in the equilibrium point is 
perpendicular to the $\mu$ axis. For every fixed value of $\mu$ in appropriate
ranges, we obtain an equilibrium configuration.  Therefore, if we call $q_i^0$
this  configuration, we have 
\begin{eqnarray}
q_i^0=q_i^0(\mu). 
\label{seq} \end{eqnarray}
These points are called {\it 
level points} and, by varying the value of the parameter $\mu$, they form a
curve  called {\it linear series} (Poincar\'e, 1885). To verify the stability of
these equilibria, we restrict our attention to a  plane $\mu=\rm const$. In
statics, the condition for an equilibrium to be stable is that the potential
energy $U$ in that point has a minimum. In our case, this statement means that 
a point on the linear series (\ref{seq}) represents a stable configuration only 
if all the different vertical sections of the surface $U =$ constant across that
point are turned towards the direction of decreasing $U$. 

It is possible to
prove (see, e.g., Jeans, 1928) that every change in the stability of a system 
corresponds to an extremum of the linear series. Accordingly, the investigation
of the stability of a self-gravitating system with respect to the gravothermal
catastrophe, is accomplished locating the first critical point on a
suitably parametrized linear series. This is a sequence of equilibrium systems
with appropriate constraints insuring that the perturbation implicit in the
stability analysis, keeps fixed the relevant physical quantities of the system.
The critical value of the parameter identifies the transition from stability to
instability.

\subsection{Choice of the constraints}

In the problem we are investigating, since the distribution functions
(\ref{DF}) depend on the three parameters $A$, $\sigma$, $\gamma$ (or
$r_{\rm a}$) and on the central potential $\Psi_{\rm c}$, we  will have to fix,
in agreement with the theory, three quantities belonging to the particular
model under consideration and let a fourth quantity vary as a function of a
parameter that is able to describe such model in a unique way. Besides keeping
fixed the total mass of the system and the parameter $A$ (see Katz, 1980, for a
discussion about this constraint), the parameter $\gamma$ has been held constant
and we will investigate the stability by means of the curves of the total energy
of the system versus the  central dimensionless potential. Our results will thus
be relevant for perturbations in which $A$, $M$, $E$ and $\gamma$ remain
constant.

As for the first three quantities we thus adopt the same constraints
used in the study of stability of isotropic systems by Katz (1980). For what
concerns the fourth quantity, it is worth noting that the anisotropy parameter
$\gamma$ is only one of the  possible choices, but other quantities related to
the anisotropy of  the configurations  could be held constant along the
equilibrium  sequences. We  could not find any convincing physical criterion in
favor of one particular choice and the parameter $\gamma$ chosen has the
advantage of allowing a straghtforward implementation of the calculation.

While it would be clearly desirable that the arguments supporting the
choices of the constraints (in particular the choice of keeping $A$ fixed) were
substantiated by a more rigorous theory on the statistical mechanics of the
gravitating systems, there are, however, several independent evidences 
supporting the validity of the basis of the present approach. In fact, in the
case of isotropic King models, the results of some Fokker-Planck simulations
(which are free of any assumption of the kind made in our investigation) by
Wiyanto et al. (1985) have confirmed with good accuracy the instability limit
found by Katz (1980) and the Fokker-Planck simulations by Cohn (1980)
have shown that, for Plummer models also, the value of the central
dimensionless potential at the onset of gravothermal instability is similar to
that found by Katz. Moreover, it is worth emphasizing that the choice of the
third parameter to be kept constant along the series is likely not to be so
important for the  conclusions one can  draw by means of this  analysis. In
fact, for example, if in the analysis of the stability of isotropic King models,
instead of keeping fixed $A$, another quantity like the tidal radius were kept
fixed, the critical value of the central potential would not change
significantly,  being in this case $\tilde \Psi_{\rm c} \simeq 8$ (see, e.g.,
the plot of $E/(G M^2 / r_{\rm t})$ versus $W_0 \equiv \tilde \Psi_{\rm c}$ in
Chernoff et al. (1986) from which this critical 
value can be obtained). Analogous results have been obtained in a more recent
analysis by Lagoute \& Longaretti (1996) who, with the same approach adopted in
the present work and the same choice of the constraints, have investigated the
gravothermal instability of rotating King models.

\begin{figure}
\vspace{8cm}  
\includegraphics{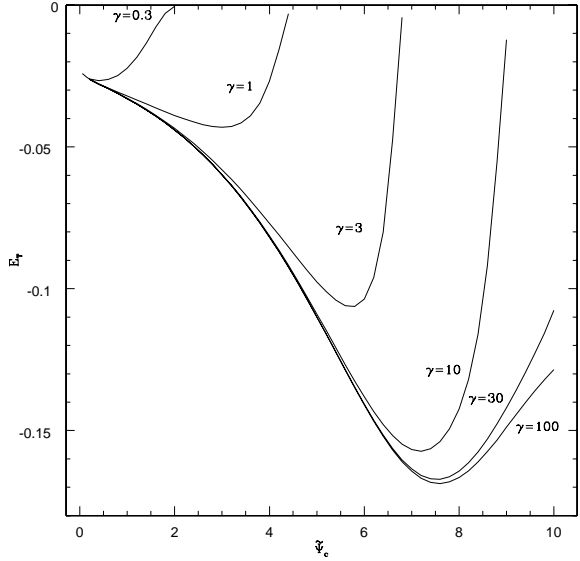} 
\caption{ E vs $\tilde \Psi_{\rm c}$ for Woolley-Dickens models.
\label{fig:fig2} }
\end{figure}
\noindent

\begin{figure}
\vspace{8cm}  
\includegraphics{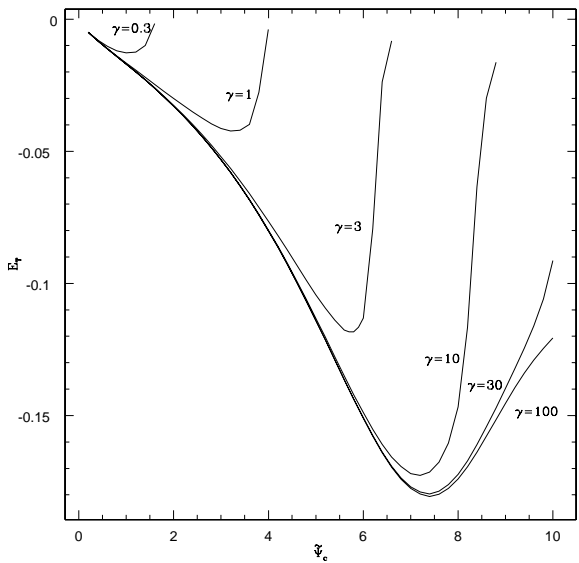}
\caption{ E vs $\tilde \Psi_{\rm c}$ for King models.
\label{fig:fig3} }
\end{figure}
\noindent

\begin{figure}
\vspace{8cm}  
\includegraphics{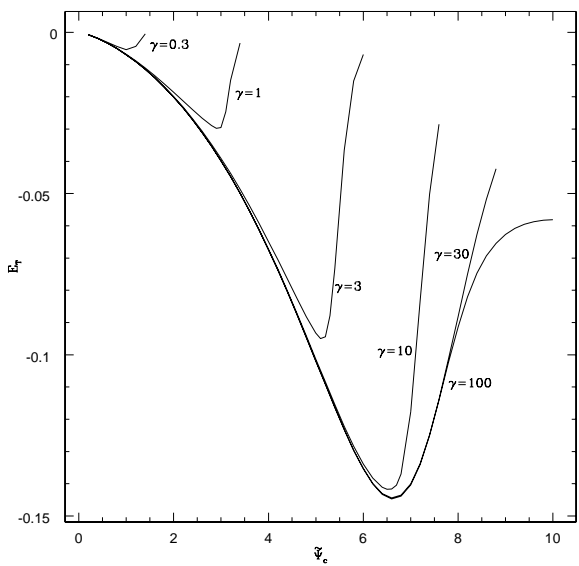}
\caption{ E vs $\tilde \Psi_{\rm c}$ for Wilson models.
\label{fig:fig4} }
\end{figure}
\noindent

\subsection{Transition to instability}

Figg. 2--4  show the plots of $E$  versus $\tilde{\Psi}_{\rm c}$ for different
values of $\gamma$, for  Woolley-Dickens, King end
Wilson models respectively. The anisotropy parameter ranges from 100, which
corresponds to nearly isotropic models, to 0.3, which corresponds to models with
a strong anisotropy. The minima of these curves show the configurations in
correspondence of which we have the transition between stable (to the left of
the minimum) and unstable models (to the right of the minimum).  

For large values of $\gamma$, only the very outer regions of the
systems are affected by the presence of anisotropy and thus the
equilibria are very similar to those of the corresponding isotropic
systems. As expected, the critical values of $\tilde \Psi_{\rm c}$ tends to
the critical value for isotropic systems listed in (\ref{cc}). As $\gamma$
decreases, the anisotropy extends deeper and deeper into the core, changing
significantly the properties of the equilibria so that the
critical value of the central potential is quite different from that of
isotropic systems. This is clearly shown in Fig. 5 in which  we have
plotted the critical values of $\tilde \Psi_{\rm c}$ versus $\gamma$ for the
models investigated. The interesting point to note is that the main effect of
anisotropy is that of causing the system to become unstable at {\it lower
values} of $\tilde \Psi_{\rm c}$.

However, we have to remark that systems with different values of $\gamma$
and the same value of $\tilde \Psi_{\rm c}$ do not have the same
structure. This means that, although Fig. 5 gives an interesting result
on the dependence of the critical value of the central potential on
the amount of anisotropy in the system, it does not provide quantitative 
information on the issue of the change of the spatial structure of the
critical model as a function of the parameter $\gamma$. In order to
investigate this point, we have focussed our attention on the
concentration parameter 
\begin{eqnarray}
c=\log \left({{r_{\rm t}} \over {r_{\rm K}}} \right),
\end{eqnarray} 
and in Fig. 6  we have plotted the concentrations
of the critical models, $c_{\rm crit}$, versus the parameter $\gamma$.

We see that, for King and Woolley-Dickens models,
$c_{\rm crit}$ is approximately constant over a wide range of values of
$\gamma$ and it starts decreasing only for $\log \gamma<0.3$, that is
for systems in which the very inner regions also are significantly
affected by anisotropy. This means that, for King and Woolley-Dickens
models, {\it the structure of the critical models is approximately independent
on $\gamma$} and the conditions for the onset of gravothermal collapse
are not significantly affected by the amount of anisotropy in the
system; {\it only for family of systems characterized by extreme  anisotropy,
deeply penetrating into the inner parts, the structure of the critical model
changes and the instability sets in for lower values of $c$}. 

The situation is different for Wilson models: for these systems
$c_{\rm crit}$ has a {\it maximum} for $\log \gamma \sim 0.5$ and then decreases
for $\log \gamma < 0.5$; in this case also, for most values of
$\gamma$ investigated, the spatial structure of the critical
equilibria does not vary much with the anisotropy.

\begin{figure}
\vspace{8cm}  
\includegraphics{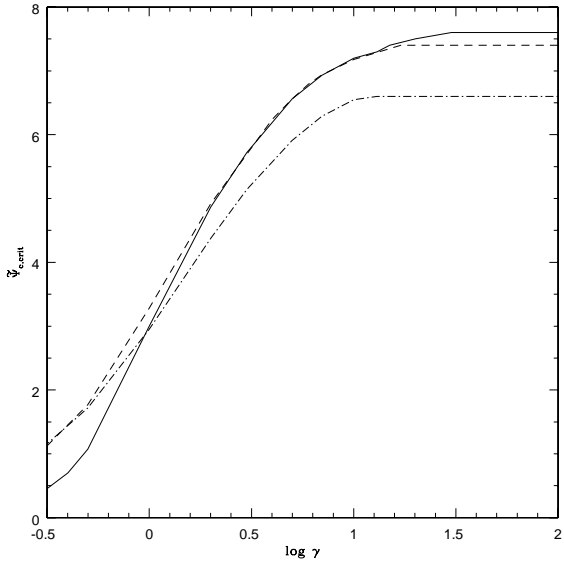} 
\caption{ Critical value for the onset of gravothermal instability of the
dimensionless central potential
 versus $\gamma$ for King (dashed line), Woolley-Dickens (solid line)
 and Wilson  (dot-dashed line) models. For any value of $\gamma$,
 models with $\tilde \Psi_{\rm c}$ larger than the critical value are
 unstable against gravothermal collapse.
\label{fig:fig5} }
\end{figure}

\begin{figure}
\vspace{8cm}  
\includegraphics{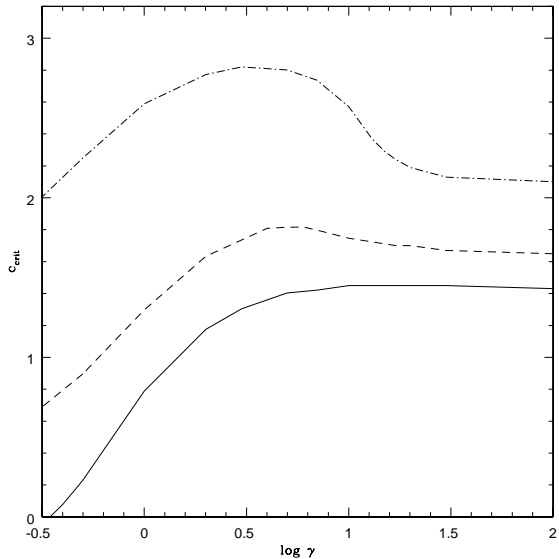} 
\caption{ Critical value (for the onset of gravothermal instability) 
 of the concentration parameter, $c$, 
 versus $\gamma$ for King (dashed line), Woolley-Dickens (solid line)
 and Wilson  (dot-dashed line) models. For any value of $\gamma$,
 models with $c$ larger than the critical value are
 unstable against gravothermal collapse.
\label{fig:fig6} }
\end{figure}

While an exact comparison  with the results of numerical
simulations of anisotropic systems (see the papers quoted in the
Introduction) is not easy, we point out that our results are in
good qualitative agreement  with the results  obtained by integration
of the Fokker-Planck equation by Takahashi (1996) and Takahashi et al.
(1997). As shown in these works, in fact, the main difference in the
evolution towards core collapse of isotropic and anisotropic systems is
in the evolutionary timescale, the former evolving faster than the
latter. There is, instead, no significant difference in the structures
of isotropic and anisotropic systems, when compared at the same
evolutionary phase (see Fig. 6 in Takahashi (1995) and Fig. 2b in
Takahashi et al. (1997)). As for the work of Takahashi et al. (1997), it
is important to note that the maximum amount of anisotropy is reached
at the moment of maximum contraction at the end of the process of core
collapse and the best-fitting anisotropic King model at that phase has
$\gamma=300$. We point out that our critical models refer to the
structure of the systems at an earlier phase than that of maximum
contraction (our critical models are meant to describe the system just at
the moment of onset of gravothermal instability), when the value of
$\gamma$ in the simulations of Takahashi et al. must be even larger. 
This means that the onset of gravothermal instability in the
simulations of Takahashi et al. occurs well in the regime of high
values of $\gamma$, where our analysis shows that no difference is
expected between the spatial structure of the critical models for the
isotropic and anisotropic case.

Additional simulations of systems with a stronger anisotropy at the
moment of the onset of core collapse would be extremely interesting in order to
test our conclusions about the decrease of the critical value of $c$
for small values of $\gamma$.

\section {Conclusions}

Models based on energy truncated distribution functions are still a very useful
tool to explore  some of the properties exhibited by stellar systems when
they are in an  enviroment that can heavily influence their
characteristics: this 
is the  case of globular clusters, but can be of relevance even in the case of 
elliptical galaxies in rich clusters.

In this paper, after having described their main structural properties, we have
investigated, by the method of linear series, the stability against gravothermal
collapse of models described by King, Wilson and Woolley-Dickens
anisotropic distribution functions. The amount of anisotropy of  a system has
been quantified simply by means of the ratio, $\gamma=r_{\rm a}/r_{\rm K}$, of
the anisotropy radius $r_{\rm a}$, which is the radius beyond which radial orbits
dominate, and the King core radius $r_{\rm K}$.

We have shown that, as $\gamma$ decreases (more anisotropic systems),
the central potential of the critical model decreases. The spatial
structure of the critical model, globally described by the concentration
parameter $c=\log (r_{\rm t} / r_{\rm K})$, remains unchanged for a wide range of
values of $\gamma$ (this is true in particular for King and Woolley-Dickens
models, while the range is smaller for Wilson models); in fact
the concentration of the critical models as a function of
$\gamma$ is approximately constant and equal to the typical value for
isotropic systems down to $\log \gamma >0.3$ ($\log \gamma>1$ for
Wilson models). Only for very anisotropic systems, with $\log
\gamma <0.3$, the concentration of the critical models decreases as
$\gamma$ decreases and the anisotropy reaches the very central regions
of the systems.

We have shown that our conclusion for  high values of
$\gamma$ is consistent  with the results of a recent investigation
carried out by means of the integration of the Fokker-Planck equation for
anisotropic systems by Takahashi et al. (1997), showing that no
significant difference exist between the spatial structure of isotropic
and anisotropic systems when they are compared at the same
evolutionary phase. 

We point out again that numerical 
simulations designed to follow the evolution of systems characterized by a
stronger anisotropy {\it at the moment} of the onset of core collapse would be
suitable to verify our conclusions 
concerning the decrease of the critical concentration for smaller
values of $\gamma$.

\vspace{1cm}
\noindent
{\bf ACKNOWLEDGEMENTS}\\
MM acknowledges support from the Isaac Newton Scholarship.

\section*{References}
Antonov V. A., 1962, Vest. Leningr. Univ., 7, 135\\
Bettwieser E., Spurzem R., 1986, A,A, 161, 102\\
Binney J., Tremaine S., 1987, Galactic Dynamics, Princeton University Press\\
Chernoff D., Kochanek C., Shapiro S. L., 1986, ApJ, 309, 183\\
Cohn H., 1980, ApJ, 242, 765\\
Giersz M., Heggie D. C., 1994a, MNRAS, 268, 257\\
Giersz M., Heggie D. C., 1994b, MNRAS, 270, 298\\
Giersz M., Heggie D. C., 1996, MNRAS, 279, 1037\\
Giersz M., Heggie D. C., 1997, MNRAS, 286, 709\\
Giersz M., Spurzem R., 1994, MNRAS, 269, 241\\
Jeans J.H., 1928, Astronomy and Cosmogony, Cambridge
University Press\\ 
Katz J., 1978, MNRAS, 183, 765\\
Katz J., 1980, MNRAS, 190, 497\\
King I. R., 1965, AJ, 70, 376\\
Lagoute C., Longaretti P. Y., 1996, AA, 308, 441\\
Louis P. D., Spurzem R., 1991, MNRAS, 251, 408\\ 
Lynden-Bell D., 1967, MNRAS, 136,  101\\
Lynden-Bell D., Wood R., 1968, MNRAS, 138, 495\\
Merritt D., Tremaine S., Johnstone D., 1989, MNRAS, 236, 829\\
Poincar\'e H., 1885, Acta Math.,  7,  259\\
Spurzem R., 1991, MNRAS, 252, 177\\
Spurzem R., Takahashi K., 1995, MNRAS, 272, 772\\
Spurzem R., Aarseth S.J., 1996, MNRAS, 282, 19\\
Stiavelli M., Bertin G., 1985, MNRAS, 217, 735\\
Takahashi K., 1995, PASJ, 47, 561\\
Takahashi K., 1996, PASJ, 48, 691\\
Takahashi K., Lee H.M., Inagaki S., 1997, MNRAS, 292, 331\\
Tremaine S. D., 1986, in  {\it Structure and Dynamics of Elliptical
Galaxies}, IAU Symp. 127, T. de Zeeuw ed.\\
Wilson C.P., 1975, AJ, 80, 175\\
Wiyanto P., Kato S., Inagaki S., 1985, PASJ, 37, 715\\
Woolley R., Dickens R. J., 1961, Royal Greenwich Obs. Bulletin, N. 42\\
Woolley R., Robertson D. A., 1956, MNRAS, 116, 288

\end{document}